# Ultra-hard hexagonal $C_{12}$ with *C3* cyclopropane-like moiety from first principles.

Samir F Matar[*]

Lebanese German University (LGU), Sahel Alma, Jounieh, Lebanon

https://orcid.org/0000-0001-5419-358X

[*]Email: s.matar@lgu.edu.lb.

**Abstract**

A novel carbon allotrope, hexagonal $C_{12}$, is proposed from crystal chemistry and quantum density functional theory DFT calculations of ground state and physical properties. The structure exhibits corner sharing distorted tetrahedra with the presence of $C_3$ triangular cyclopropane-like moiety connecting planar carbon. $C_{12}$ allotrope is found cohesive and stable both mechanically (elastic constants and their combinations) and dynamically (phonons band structures) and presents ultra-hardness with $H_{Vickers}$ = 70 GPa. The temperature dependence of the heat capacity $C_V$ shows close magnitudes to experimental results of diamond. The electronic band structure reveals insulating behavior with 2.5 eV band gap, i.e., half smaller than diamond's.

**Introduction**

In the field of the challenging search for ultra-hard materials, carbon occupies a particular position especially with diamond recognized as the hardest material. The tetrahedral $C(sp^3)$ hybridization makes diamond a perfectly covalent electronic system with large electronic band gap of 5 eV. A change to semi-conductive to metallic behaviors for applications in electronics can be induced by self-doping with $C(sp^2)$ as shown recently in extended carbon tetrahedral networks (cf. [1] and therein cited works), as well as with structural strains arising from distorted *C4* tetrahedra as shown in present work.

In 2016 Cheng et al. [2] proposed 3D (three dimensional) metallic Tri-$C_9$ built by distorting sp$^3$ within a structure made of *$C_3$* cyclopropane-like moieties; a feasible synthesis route was then proposed by compressing graphite. In the rhombohedral structure the representative distorted tetrahedron shown in Fig. 1 exhibits angles departing from the ideal 109.4° value with expectedly 60° in the cyclopropane unit and 123° on the opposing side. An earlier work proposed $C_{12}$ characterized by $C(sp^2)$ in a developed study of carbon-nanotube polymers with cyclopropane moiety based on ab initio particle-swarm optimization [3]. Note that *$C_3$* moiety was known back in 2010 to take part in the hexagonal structure of (hypothetic) "Trihexagonite" $C_{10}$ [4] characterized by two graphitic-like layers connected by *$C_3$* moiety as shown in Fig. 1b.

In the context of carbon allotropes with *$C_3$* cyclopropane moiety, a novel 3D hexagonal carbon allotrope $C_{12}$ is proposed based on quantum mechanics investigations within the well-established density functional theory (DFT) [5,6]. The structure shown in Fig. 1c was obtained after unconstrained geometry optimization onto the ground state, *vide infra*. It is characterized by distorted tetrahedra as in Figs. 1a and 1b and demonstrated to be cohesive with both mechanical and dynamic stabilities. $C_{12}$ is identified as ultra-hard and possessing an electronic band gap half smaller than diamond, as well as similar thermal properties to diamond.

**1- Computational framework**

The *modus operandi* consisted in submitting the devised structures to unconstrained geometry relaxations of the atomic positions and the lattice constants down to the ground state characterized by minimum energy and subsequent cohesive energy. The iterative computations were performed using DFT-based plane-wave Vienna Ab initio Simulation Package (VASP) [7,8]. For the atomic potentials, the projector augmented wave (PAW) method was used [8,9]. The exchange X and correlation effects (XC) were treated within a generalized gradient approximation scheme (GGA) following Perdew, Burke and Ernzerhof PBE [10]. Using screened Coulomb potential-based hybrid functionals as HSE06 [11] in trial preliminary calculations did not lead to better results versus GGA-PBE. The relaxation of the atoms onto ground state geometry was done applying a conjugate-gradient algorithm [12]. Blöchl tetrahedron method [13] with corrections according to Methfessel and Paxton scheme [14] were applied for geometry optimization and energy calculations respectively. A special *k*-point sampling [15] was applied for approximating the reciprocal space Brillouin-zone (BZ) integrals. For better reliability, the optimization of the structural parameters was carried out along with successive self-consistent cycles with increasing k-mesh until the forces on atoms were less than 0.02 eV/Å and the stress components below 0.003 eV/Å$^3$.

Mechanical stability and hardness were obtained from the calculations of the elastic constants. The treatment of the results was done thanks to ELATE [16] online tool devoted to the analysis of the elastic tensors providing the bulk B and shear G moduli along different averaging methods; Voigt's based on a uniform strain was used herein. For the calculation of the Vickers hardness a semi-empirical model based on elastic properties was used [17].

The phonon frequencies were calculated to verify the dynamic stability. The corresponding modes were computed considering the harmonic approximation through finite displacements of the atoms around their equilibrium positions to obtain the forces from the summation over the different configurations. The phonon band structures along the directions of the hexagonal Brillouin zone were subsequently obtained using "Phonopy" interface code based on Python language [18]. The structure sketches including the tetrahedral representations were produced with VESTA graphic software [19]. The electronic band structures were obtained with the full-potential augmented spherical wave ASW method [20] based on DFT using the same GGA scheme as in above protocol for ground state search.

   2- **Crystal chemistry**

Hexagonal $C_{10}$ (Fig. 1b) crystallizes in space group *P*-6*m*2 (N° 187) with a topology $3^2.4^2$**T**201 describing the underlying net as identified from Topcryst program [21]. The structure used herein is catalogued in SACADA carbon database [22] at N°206. Table 1 shows the literature [4] and presently calculated (in brackets) lattice constants and internal coordinates after full unconstrained geometry optimizations. A good agreement can be found. This also stands for the interatomic distances, especially the smallest one within the $C_3$ moiety, i.e., the distinguishing constituent of the structure. Note that the $C_{10}$ structure features a single carbon C at the origin (0, 0, 0), and 9 C at three 3-fold Wyckoff positions. Comparatively, the major difference presented with devised $C_{12}$ is in the 4 three-fold Wyckoff position at 2 (3j) and 2 (3k). The fully converged parameters and internal coordinates are given in Table 1. The same space group *P*-6*m*2 (N° 187) was assigned. The interatomic distances are now scattered over a larger range of magnitudes featuring small C-C separations. Regarding the density $\rho(C_{10})$ =3.09 g/cm$^3$ and $\rho(C_{12})$ =3.32 g/cm$^3$; the difference arising from relatively smaller distances in the latter. Note the highest density for 3D carbon is observed for diamond with $\rho$ =3.56 g/cm$^3$. Subsequently, $C_{12}$ was submitted to TopCryst analysis revealing zeolite-related **bba** topology, like C(sp$^2$) $C_{12}$ also containing 'tripropanoïd' *C3* moiety [3] (SACADA Database Code: 118, Topology: **bba**-3,4; *P*6/*mmm*), i.e., different from $C_{10}$ (3$^2$.4$^2$**T**201 topology). Energetically, $C_{12}$ was found significantly more cohesive than $C_{10}$ as shown in the last line of Table 1, while remaining less cohesive than diamond: $E_{coh.}$ = -2.49 eV/at. (cf. [1]).

### 3- Mechanical properties from elastic constants

The analysis of the mechanical behavior was subsequently carried out with the elastic properties. The calculated sets of elastic constants $C_{ij}$ (i and j correspond to directions) obtained by performing finite distortions of the lattice are given in Table 2. For the sake of comparison, the elastic constants were calculated for $C_{10}$ besides $C_{12}$. All $C_{ij}$ values are positive signaling stability of the two systems. Using ELATE [16] introduced above, the bulk (*B*) and the shear (*G*) moduli obtained by averaging the elastic constants using Voigt's method are given in Table 2. The hardness magnitudes calculated with Chen et al. method use the elastic properties are given at the last column of Table 2.

$C_{12}$ has the largest values with $B_V$ =409 GPa and $G_V$= 421 GPa that remain smaller than the accepted values for diamond, $B_V$ =445 GPa and $G_V$ = 550 GPa [23]. The corresponding Vickers hardness ($H_V$) magnitudes clearly show the larger hardness of $C_{12}$ expected from the higher density versus $C_{10}$. The difference between the two allotropes is the purely tetrahedral structure of presently devised system whereas trigonal carbon characterized $C_{10}$. Such differences are numerically featured with the Pugh G/V ratio larger than 1 for $C_{12}$ indicating a trend to brittleness versus G/V<1 for $C_{10}$ indicating a trend to ductility.

After these comparison results we focus on novel $C_{12}$ for the study of the dynamic, thermal, and electronic band structure in the second part of the paper.

### 4- Dynamic properties from the phonons

An important criterion of phase stability is obtained from the phonon's properties. Phonons are quanta of vibrations; their energy is quantized by the Planck constant 'h' which is used in its reduced form ℏ (ℏ = h/2π), giving the phonons energy: E = ℏω (frequency: ω)

Fig. 2 shows the phonon bands. In the horizontal direction, the bands develop along the main lines of the hexagonal Brillouin zone (reciprocal k-space). The vertical direction shows the frequencies ω given in terahertz (THz) units.

There are 3N-3 optical modes found at higher energy than three acoustic modes that start from zero energy (ω = 0) at the Γ point, center of the Brillouin zone (BZ), up to a few terahertz. They correspond to the lattice rigid translation modes of the crystal (two transverse and one longitudinal). The remaining bands correspond to the optical modes and culminating at ω ~ 45 THz.

Complementary molecular quantum chemistry calculations of cyclopropane $C_3H_6$ showed that this frequency equivalent of 1510 cm$^{-1}$ corresponds to the beathing mode of the *C3* cyclopropane trangle, only active in Raman. The second highest frequency is found at 39 THz close to the value observed for diamond of 40 THz also by Raman spectroscopy [24].

## 5- Thermal properties

The thermodynamic properties such as entropy S and heat capacity at constant volume (Cv) were then calculated from the phonon frequencies using statistical thermodynamics [25]. Fig. 3 shows the change with temperature of the calculated entropy and heat capacity of $C_{12}$ in comparison with corresponding experimental data for diamond [26]. Expectedly, entropy S increases with temperature. Heat capacity Cv increases following a nonlinear evolution with a trend to flatten at highest temperature. This behavior is close to the experimental values of diamond that scatter over the calculated $C_{12}$ curve. It can be suggested that $C_{12}$ is close to diamond at least thermally. Thermal conductivity details are provided in a recent work [27].

Knowing that diamond is a wide band gap electronic insulator, we calculated the electronic band structure of $C_{12.}$

## 6- Electronic band structures and density of states

Using the crystal parameters in Table 1, the electronic band structure was obtained using the all-electrons DFT-based augmented spherical method (ASW) [20]. In Figure 4 the bands develop along the main directions of the primitive hexagonal Brillouin zone. Along the energy y-axis the energy zero is taken with respect to the top of the valence band VB separated from the conduction band CB by a band gap amounting to ~2.5 eV signaling an insulating system alike diamond but with half magnitude band gap that can be assigned to the structure made of distorted tetrahedra. Lastly we note that $C_{10}$ was found with metallic behavior as mentioned in the literature.

**Conclusions**

Novel carbon allotrope $C_{12}$ has been proposed based on crystal chemistry and quantitative assessments of the physical properties within quantum mechanical density functional theory (DFT). The novel allotrope characterized by *C3* cyclopropane-like moiety belongs to a family of similar *C3*-containing allotropes with which comparisons were established. With a distorted tetrahedral network, $C_{12}$ was found cohesive, and mechanically stable (elastic properties) characterized by ultra large hardness (just below diamond) and stable dynamically from the computed phonons. Furthermore, the thermal properties reporting on the specific heat were found close to diamond experimental values. The electronic band structure indicates smaller band gap than diamond that could be assigned to the distorted tetrahedral structure.

REFERENCES

[1] S.F., Matar, V. Eyert, V.L. Solozhenko V.L. Novel Ultrahard Extended Hexagonal $C_{10}$, $C_{14}$ and $C_{18}$ Allotropes with Mixed $sp^2$/$sp^3$ Hybridizations: Crystal Chemistry and Ab Initio Investigations. *C*, **9**, (2023), 11. https://doi.org/10.3390/c9010011

[2] Y. Cheng, R. Melnik, Y. Kawazoe, B. Wen B. Three-Dimensional Metallic Carbon from Distorting $sp^3$-Bond. *Cryst. Growth Des*. **16** (2016), 3, 1360–1365. https://doi.org/10.1021/acs.cgd.5b01490

[3] Zhisheng Zhao, Bo Xu, Li-Min Wang, Xiang-Feng Zhou, Julong He, Zhongyuan Liu, Hui-Tian Wang, and Yongjun Tian. Three-Dimensional Carbon-Nanotube Polymers. *ACS Nano*, **5** (2011), 7226–7234.

[4] M.J. Bucknum, B. Wen, E.A Castro. Trigohexagonite. *Journal of Mathematical Chemistry*. **48**, (2010), 816–826

[5] P. Hohenberg, W. Kohn, Inhomogeneous electron gas. *Phys. Rev. B* **136** (1964) 864-871.

[6] W. Kohn, L.J Sham, Self-consistent equations including exchange and correlation effects. *Phys. Rev. A* **140** (1965) 1133-1138.

[7] G. Kresse, J. Furthmüller, Efficient iterative schemes for ab initio total-energy calculations using a plane-wave basis set. *Phys. Rev. B* **54** (1996) 11169.

[8] G. Kresse, J. Joubert, From ultrasoft pseudopotentials to the projector augmented wave. *Phys. Rev. B* **59** (1994) 1758-1775.

[9] P.E. Blöchl, Projector augmented wave method. *Phys. Rev. B* 50 17953-17979.

[10] J. Perdew, K. Burke, M. Ernzerhof, The Generalized Gradient Approximation made simple. *Phys. Rev. Lett.* **77** (1996) 3865-3868.

[11] J. Heyd, G.E. Scuseria, M. Ernzerhof, Hybrid functionals based on a screened Coulomb potential. *J. Chem. Phys.* **124** (2006) 219906.

[12] W.H. Press, B.P. Flannery, S.A. Teukolsky, W.T. Vetterling, *Numerical Recipes*, 2$^{nd}$ ed., Cambridge University Press: New York, USA, 1986.

[13] P.E. Blöchl, O. Jepsen, O.K. Anderson, Improved tetrahedron method for Brillouin-zone integrations. *Phys. Rev. B* **49** (1994) 16223-16233.

[14] M. Methfessel, A.T. Paxton, High-precision sampling for Brillouin-zone integration in metals. *Phys. Rev. B* **40** (1989) 3616-3621.

[15] H.J. Monkhorst, J.D. Pack, Special k-points for Brillouin Zone integration. *Phys. Rev. B* **13** (1976) 5188-5192.

[16] R. Gaillac et al. ELATE: an open-source online application for analysis and visualization of elastic tensors. *J. Phys.: Condens. Matter* **28** (2016) 275201.

[17] X.Q. Chen, H. Niu, D. Li, Y. Li, Modeling hardness of polycrystalline materials and bulk metallic glasses. *Intermetallics* **19** (2011) 1275-1281.

[18] A. Togo, I. Tanaka, First principles phonon calculations in materials science. *Scr. Mater*. **108** (2015) 1-5.

[19] K. Momma, F. Izumi, VESTA 3 for three-dimensional visualization of crystal, volumetric and morphology data. *J. Appl. Crystallogr*. **44** (2011) 1272-1276.

[20] V. Eyert, Basic notions and applications of the augmented spherical wave method. *Int. J. Quantum Chem.* **77** (2000) 1007-1031.

[21] A.P. Shevchenko, A.A. Shabalin, I.Yu. Karpukhin, V.A. Blatov, Topological representations of crystal structures: generation, analysis and implementation in the TopCryst system. *Sci Technol Adv Mat*. **2** (2022) 250-265.

[22] SACADA database (*Samara Carbon Allotrope Database*). www.sacada.info

[23] V.V. Brazhkin, V.L. Solozhenko, Myths about new ultrahard phases: Why materials that are significantly superior to diamond in elastic moduli and hardness are impossible. *J. Appl. Phys*. **125** (2019) 130901.

[24] R.S. Krishnan, Raman spectrum of diamond, *Nature* **155** (1945) 171.

[25] M.T. Dove. Introduction to Lattice Dynamics, Cambridge University Press, 1993. 29

[26] Victor A.C. Heat capacity of diamond at high temperatures, *J. Chem. Phys*. (1962) **36** 1903–1911.

[27] Ch. Liu, Y. Chen, Ch. Dames. Electric-Field-Controlled Thermal Switch in Ferroelectric Materials Using First-Principles Calculations and Domain-Wall Engineering. *Phys. Rev. Applied* (2019) 11, 044002.

TABLES

Table 1. Crystal parameters of the hexagonal C allotropes.

| P-6m2 N°187 | C$_{10}$ ([4]) | C$_{12}$ ( pw ) |
|---|---|---|
| a, Å | 5.3410 (5.3642) | 5.771 |
| c, Å | 2.6056 (2.5804) | 2.499 |
|  | C1 (1a) 0.0, 0.0, 0.0 | C1 (3j) 0.3471, 0.1736, 0.0 |
|  | C2 (3j) 0.3278 (0.3275), 0.1639 (0.1637), 0.0 | C2 (3j) 0.5069, 0.0138, 0.0 |
|  | C3 (3j) 0.5075 (0.5072), 0.0151 (0.0143) 0 | C3 (3k) 0.4248, 0.8409, ½ |
|  | C4 (3k) 0.4238 (0.4232), 0.8477 (0.8464),½ | C4 (3k) 0.1743, 0.0872, ½ |
| Shortest distances, Å | C4-C4= 1.447 (1.450) | C3-C3 =1.509 |
|  | C1-C2 = 1.521 Å | C3-C2 = 1.519 |
|  | C3-C2 = 1.531 | C1-C4=1.519 |
|  |  | C1-C2 = 1.597 |
| Density ρ (g/cm$^3$) | 3.09 | 3.32 |
| Cohesive Energy, eV | -1.85 | -2.13 |

pw = present work

Table 2. Elastic constants C$_{ij}$ and Voigt values of bulk ($B_V$) and shear ($G_V$) moduli and Vickers hardness of carbon allotropes (all values are in GPa).

|  | C$_{11}$ | C$_{12}$ | C$_{13}$ | C$_{33}$ | C$_{44}$ | C$_{66}$ | B$_{Voigt}$ | G$_{Voigt}$ | H$_{Vickers}$ |
|---|---|---|---|---|---|---|---|---|---|
| C$_{10}$ | 942 | 176 | 81 | 737 | 383 | 178 | 366 | 300 | 41 |
| C$_{12}$ | 936 | 172 | 43 | 1210 | 382 | 392 | 409 | 427 | 70 |

FIGURES

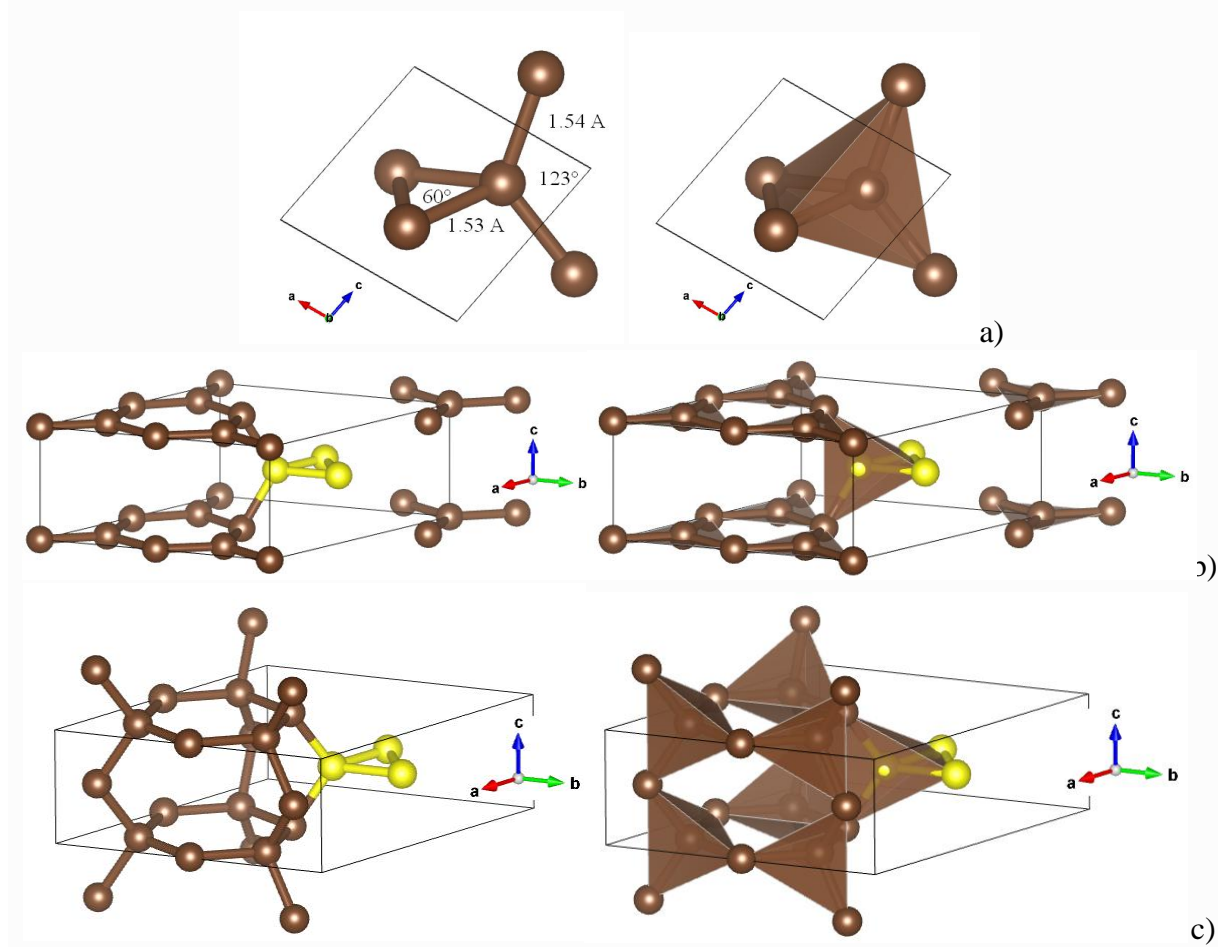

Figure 1. a) Distorted C(sp$^3$) building block in Tri-C$_9$ (rhombohedral setting) [2], b) C$_{10}$ [4], and c) C$_{12}$ (present work). Structures are shown in ball-and-stick and tetrahedral representations.

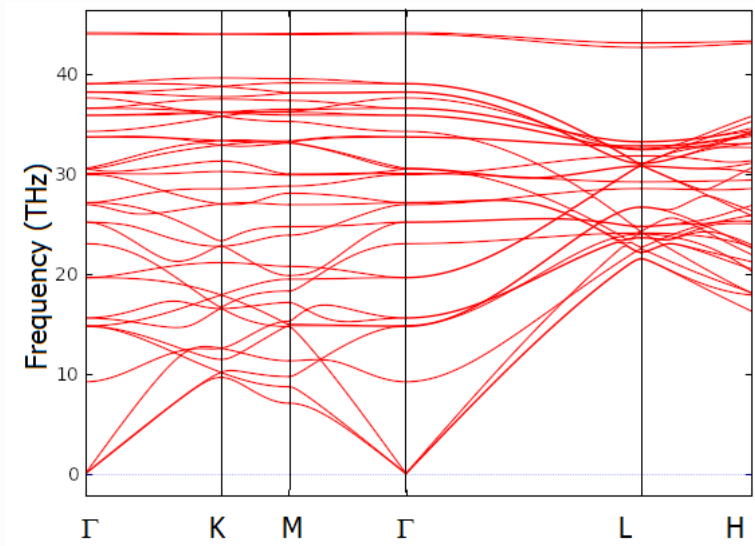

Figure 2. Phonons band structures of $C_{12}$

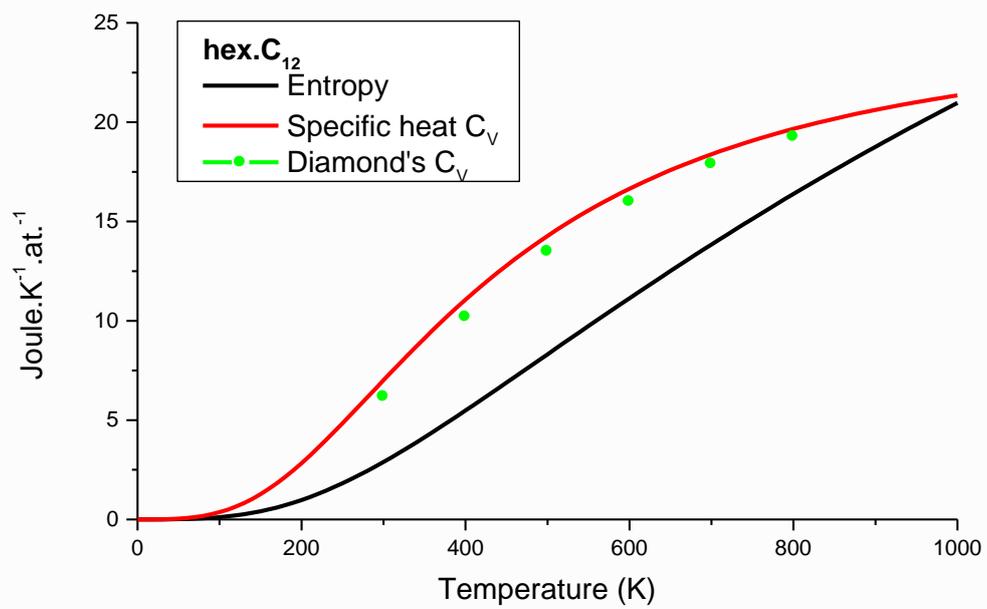

Figure 3. Thermal properties of $C_{12}$ compared to diamond for the specific heat $C_V$.

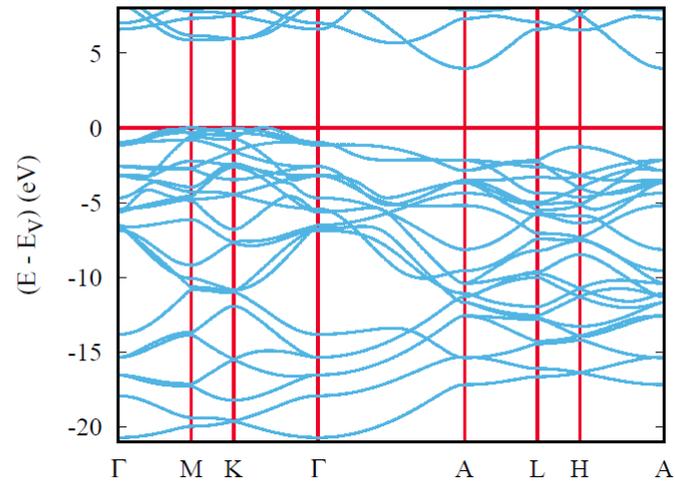

Figure 4. Electronic band structure of $C_{12}$